\documentclass[aps,pre,twocolumn,superscriptaddress,superscriptreference]{revtex4-2}

\usepackage{amsmath,bbold,bm,amssymb,scalerel,mathtools}
\usepackage{graphicx}
\usepackage{color}
\usepackage{enumitem}
\usepackage{algorithm,algpseudocode}
\usepackage{multirow}
\usepackage{colortbl,booktabs}
\usepackage{placeins}
\usepackage[usenames,dvipsnames]{xcolor}

\usepackage[colorlinks, linkcolor=BrickRed, urlcolor=blue!50!black, citecolor=blue!50!black]{hyperref}

\newcommand\equalhat{%
	\let\savearraystretch\arraystretch
	\renewcommand\arraystretch{0.3}
	\begin{array}{c}
		\stretchto{
			\scalerel*[\widthof{=}]{\wedge}
			{\rule{1ex}{3ex}}%
		}{0.5ex}\\ 
		=%
	\end{array}
	\let\arraystretch\savearraystretch
}
\newcommand{\myrowcolour}{\rowcolor[gray]{0.925}}

\usepackage[normalem]{ulem}

\begin{document}

\title{Non-Markovian systems out of equilibrium: Exact results for two routes of coarse graining}

\author{Gerhard Jung}
\email{jung.gerhard@umontpellier.fr}
\affiliation{ Department of Chemical Engineering, Kyoto University, Japan }
\affiliation{Laboratoire Charles Coulomb (L2C), Université de Montpellier, CNRS, 34095 Montpellier, France}

\begin{abstract}
Generalized Langevin equations (GLEs) can be systematically derived via dimensional reduction from high-dimensional microscopic systems. For linear models the derivation can either be based on projection operator techniques such as the Mori-Zwanzig (MZ) formalism or by ``integrating out'' the bath degrees of freedom. Based on exact analytical results we show that both routes can lead to fundamentally different GLEs and that the origin of these differences is based inherently on the non-equilibrium nature of the microscopic stochastic model. The most important conceptional difference between the two routes is that the MZ result intrinsically fulfills the generalized second fluctuation-dissipation theorem  while the integration result can lead to its violation. We supplement our theoretical findings with numerical and simulation results for two popular non-equilibrium systems: Time-delayed feedback control and the active Ornstein-Uhlenbeck process.

\end{abstract}

\maketitle

\section{Introduction}

Coarse graining describes the process where a high-dimensional microscopic system is replaced by a mesoscopic model with much fewer degrees of freedom \cite{muller2002coarse,izvekov2005multiscale,peter2009multiscale,klippenstein2021introducing,schilling2021coarse}. These coarse-grained (CG) models can be computationally much more efficient than the original system, thus they allow bridging the gaps between the microscopic high-dimensional microscopic and the macroscopic time and length scales. Consequently, coarse graining has become an integral part of statistical physics and computer simulations of soft matter systems \cite{voth2008coarse,brini2013systematic}. A fundamental problem behind such a CG procedure is, however, that the resulting equations of motion typically become non-Markovian. Using the Mori-Zwanzig projection operator (MZ) formalism it has been shown that the CG dynamics for an observable $ A(t) $ can be described by a generalized Langevin equation (GLE), including a time-dependent friction kernel, $ K(t) $, which quantifies these non-Markovian dynamics \cite{Zwanzig1961,Mori1965,Zwanzig2001} \footnote{In this work we will purely focus on one-dimensional motion.}. In equilibrium, this friction kernel is uniquely connected to the thermal fluctuations, $\eta(t)$, via the generalized second fluctuation-dissipation theorem (2FDT) \cite{Zwanzig2001},
\begin{equation}\label{eq:2FDT}
\left \langle \eta(t) \eta(0)  \right \rangle_\text{eq} = \left \langle A(0)^2 \right \rangle_\text{eq} K(t).
\end{equation}
Furthermore one can show in equilibrium that the 2FDT is a direct consequence of the first fluctuation-dissipation theorem, which connects time-correlation functions with the time-dependent dissipative response to a small perturbation \cite{Onsager1931A,Onsager1931B,kubo1966fluctuation}. Using the above relations, many coarse-graining techniques have been suggested which explicitly incorporate equilibrium non-Markovian dynamics into the equations of motion for the mesoscopic coarse-grained model \cite{li2015incorporation,li2016comparative,lei2016data,jung2017iterative,jung2018generalized,wang2020data,D1SM00413A}.

The situation is much less clear in non-equilibrium steady-states and seemingly contradictory results exist that either report validity of the 2FDT using arguments based on the MZ formalism \cite{doi:10.1063/1.5006980,zhu2021effective,zhu2021generalized,jung2021fluctuation,schilling2021coarse}\footnote{As is usually done in the literature, we will refer to Eq.~(\ref{eq:2FDT}) as a \emph{theorem} even in non-equilibrium situations where it might not be valid anymore.} or its violation \cite{Maes2014,maes2014second,Zaccone2018_FDTNESS,netz2018fluctuation,PhysRevE.101.032408,PhysRevE.102.052119, doerries2021correlation}. One reason for these controversial results is that (infinitely) many pairs $ \{K(t),\left \langle \eta(t) \eta(0)  \right \rangle_\text{neq}\} $ exist for which the GLE gives the same time autocorrelation function, but only one of these pairs actually fulfills the 2FDT \cite{PhysRevE.101.032408}. The fundamental question is therefore how to decide which of these pairs yields the most suitable representation of the microscopic system? 

On the one hand, a natural choice is to use the memory kernel resulting from the MZ formalism. Under very mild assumptions it has recently been shown that the resulting GLE fulfills the 2FDT even for stochastic microscopic dynamics \cite{zhu2021generalized}. Complementary to these results, we have derived  in Ref.~\cite{jung2021fluctuation} for arbitrary (even non-stationary) microscopic dynamics a numerical construction of memory kernels from given time autocorrelation functions which similarly fulfill the 2FDT. In the following we will denote this class of memory kernels as the ``projection route''. On the other hand, for linear models such as the one analyzed in this work, it is possible to analytically derive GLEs which clearly show violation of the 2FDT in non-equilibrium situations \cite{netz2018fluctuation,PhysRevE.102.052119,doerries2021correlation}. The analytical procedure is to ``integrate out'' the bath degrees of freedom and thus derive closed equations of motion for the selected variables only. We will call these analytical solutions the ``integration route''.

The above, seemingly contradictory results, lead to the surprising conclusion that, even without non-linear conservative or external forces \footnote{Any non-linearities in the microscopic model will lead to time-retarded memory effects in the linearized GLE, as discussed in Refs.~\cite{Zwanzig2001,10.1209/0295-5075/ac35ba}. Here, however, we start from a purely linear microscopic system.}, the projection route and the integration route lead to fundamentally different results in non-equilibrium systems. To the best of our knowledge this conclusion has not been discussed explicitly in the literature before. On the contrary, in many publications ``projecting out'' and ``integrating out'' are used as synonyms. In this work, we prove the validity of the above conclusion based on exact analytical results for a linear stochastic model. For this model it is possible to solve the projection operator formalism explicitly and to calculate analytical results for the thermal fluctuations and thus the memory kernel. We show that these expressions are fundamentally different from the previously published solutions for the integration route which have been shown to violate the 2FDT. We also perform computer simulations of the microscopic, stochastic model and show that the numerical results are in very good agreement with the analytical calculations.

Our manuscript is organized as follows. We present the microscopic stochastic model in Section \ref{sec:model} and determine exact results for the instantaneous fluctuations and the GLE via the integration route. We then introduce the MZ projection operator formalism for stochastic systems in Section \ref{sec:theory} and perform analytical calculations by explicitly evaluating the various formal expressions. In Section \ref{sec:results} we then present complementary simulation results together with numerical data for the analytical results for various non-equilibrium systems. We summarize and conclude in Section \ref{sec:conclusions}.

 \section{Microscopic Model And Exact Results}
 \label{sec:model}

The model consists of a reference particle which is described by its velocity $ v_0 $ and is referred to as colloid in the following. The colloid couples dissipatively to the velocities $ v_i $ of $ N $ other ``bath particles''. We assume that the underlying equations of motion for these particles are given by the following stochastic differential equations (SDE),
\begin{align}
m_0\dot{v}_0(t) &= -\gamma_0 v_0(t) + \sum_{i=1}^N k_i v_i(t),\label{eq:colloid}\\
m_i\dot{v}_i(t) &= - \gamma_i v_i(t) + b_i v_0 + \sqrt{2 k_B T \gamma_i } W_i(t), \quad i> 0, \label{eq:solvent}
\end{align}
with Gaussian white noise $ \langle W_i(t) W_j(t) \rangle = \delta_{ij} \delta(t) $, masses $ m_i $, friction constants $ \gamma_i $ and coupling constants $ k_i $ and $ b_i $. Throughout the manuscript we set $ m_0 = m_i = 1.0 $, $ \gamma_1 = 1.0 $ and $ k_B T = 1.0 $ which defines the units of the system. 

The above SDE can be interpreted as an already coarse-grained description of a first principle system. The friction constants $ \gamma_i $ and white noise $ W_i(t) $ then describe the coupling to a heat bath. The system is in equilibrium with temperature $ T $ for the special case of reciprocal interactions, $ b_i = -k_i $ and friction constant $ \gamma_0 =0 $ \cite{doerries2021correlation}. Here, we define as equilibrium those systems in which the equipartition theorem is fulfilled, $\langle v_i(0) v_j(0) \rangle = k_B T \delta_{ij} $. We will also consider two exemplary non-equilibrium situations. The first one is given by setting the coupling constants $ b_i = k_i $, which can lead to time-delayed friction kernels with a maximum at $ t>0 $. Such systems have recently attracted attention in the context of stochastic thermodynamics \cite{loos2014delay,loos2019fokker,Loos2020,e23060696}. The second system is defined by choosing $ b_i = 0 $, which is equivalent to the stochastic model suggested by Wu and Libchaber to describe passive colloids in active bacteria baths \cite{wu2000particle}, and is also known as active Ornstein-Uhlenbeck process \cite{PhysRevLett.117.038103}. Despite its simplicity, the microscopic SDE is therefore able to model relatively distinct non-equilibrium conditions \cite{doerries2021correlation}.

\subsection{Instantaneous fluctuations}

The linear SDE (\ref{eq:colloid}) and (\ref{eq:solvent}) can be rewritten into the matrix equation,
\begin{equation}\label{eq:lin_eom}
\dot{\bm{v}}(t) = - \mathbf{A} \bm{v}(t) + \bm{\Phi} \mathbf{W}(t),
\end{equation}
with interaction matrix $ [\mathbf{A}]_{ij} = A_{ij} $ and noise matrix $ [\bm{\Phi}]_{ij} = \Phi_{ij} $. For such a linear system one can derive an equation for the instantaneous fluctuations, $ [\mathbf{E}]_{ij} = E_{ij} = \langle v_i(0) v_j(0) \rangle $ in the steady state \cite{netz2018fluctuation},
\begin{equation}\label{eq:inst_fluct}
2 \bm{\Phi} \bm{\Phi} = \mathbf{A} \mathbf{E} + \mathbf{E}^T \mathbf{A}^T.
\end{equation}
Using SageCell \cite{sagemath} we can therefore easily derive analytical expressions for the instantaneous fluctuations $\langle v_i(0) v_j(0) \rangle $, which are essential input for the projection operator formalism. Two important relations can immediately be derived from the above analytical expressions. The first equation is,
\begin{align}
 -\gamma_0 \langle v_0(0)^2  \rangle&= - \sum_{i>0} k_i \langle v_0(0) v_i(0)  \rangle  \label{eq:rel_v01},
\end{align}
and, for the special case in which $ \gamma_0 = 0 $ and $ \gamma_i = \gamma \,\, \forall i $, it is also possible to derive the relation,
\begin{align}
\langle v_0(0)^2  \rangle &= k_B T\frac{\sum_{i>0} k_i^2 }{-\sum_{i>0}b_ik_i }. \label{eq:rel_v0}
\end{align}

 \subsection{Non-Markovian equations of motion for the colloid (integration route)}
 
Our general goal is to derive non-Markovian equations of motion for the colloid, by systematically reducing the dimensionality of the system. It has been shown in various publications for linear systems such as the one studied in this work that a coarse-grained model can be derived by integrating Eq.~(\ref{eq:solvent}) for the bath particles in time from $ 0 $ to $ t $ and thus determine the implicit solution of the SDE. This solution can then be inserted into Eq.~(\ref{eq:colloid}) \cite{Zwanzig2001,netz2018fluctuation,Loos2020,doerries2021correlation}. For the specific SDE defined above, the solution has been discussed in Ref.~\cite{doerries2021correlation},
 \begin{equation}\label{eq:GLE_integrate}
 \dot{v}_0(t) = - \gamma_0 v_0(t) - \int_{0}^{t} K^\text{I}(t-s) v_0(s) + \eta^\text{I}(t),
 \end{equation}
 with memory kernel,
 \begin{align}
 \label{eq:memory_integrate}
 K^\text{I}(t) &= -\sum_{i>0} k_i b_i \exp (-\gamma_i t),
 \end{align}
 noise 
  \begin{align}
 \label{eq:noise_integrate_full}
 \eta^\text{I}(t)&= \sqrt{2 k_B T \gamma_i } \sum_{i>0} \int_{0}^t \text{d}t^\prime k_i  \exp (-\gamma_i t^\prime) W_i(t^\prime),
 \end{align}
  and thus auto-correlation function of the noise,
 \begin{align}
\label{eq:noise_integrate}
C^\text{I}_\eta(t)&=\langle \eta^\text{I}(t) \eta^\text{I}(0) \rangle = k_B T \sum_{i>0}k_i^2  \exp (-\gamma_i t).
\end{align}
From these equations it can already be deduced that only in the specific cases in which $ \gamma_0=0 $ the 2FDT can be fulfilled. The first special case is equilibrium, i.e., $ b_i = - k_i $, for which,
\begin{align}
C^\text{I}_\eta(t)&=  - k_B T  \sum_{i>0} k_i b_i  \exp (-\gamma_i t) = \langle v_0(0)^2  \rangle K^\text{I}(t).
\end{align}
The second case is a non-equilibrium system in which $ \gamma_i = \gamma \,  \forall i>0 $, but $ b_i $ and $ k_i $ can be chosen freely,
\begin{align}
C^\text{I}_\eta(t)&= k_B T \exp (-\gamma t)\sum_{i>0} k_i^2 \nonumber \\ &= - \langle v_0(0)^2  \rangle  \exp (-\gamma t)  \sum_{i>0} k_i b_i  
= \langle v_0(0)^2  \rangle K^\text{I}(t),
\end{align}
where we have used Eq.~(\ref{eq:rel_v0}). This second case is particularly interesting because it shows that in this specific model non-equilibrium does not imply that the 2FDT is violated.

\section{Projection operator formalism}
\label{sec:theory}

Another systematic route to reduce the dimensionality of a system is given by the projection operator formalism \cite{Zwanzig1961,Mori1965,grabert2006projection,Zwanzig2001}. In the following, we will focus on the Mori-Zwanzig formalism, which was originally developed for Hamiltonian dynamics, but has been shown to also be applicable to stochastic systems \cite{PhysRevE.52.1734,10.1143/PTP.64.500,zhu2021effective,zhu2021generalized}. In the next subsection, we will shortly recap the important properties of the MZ formalism along the lines of Ref.~\cite{zhu2021generalized}, and then explicitly evaluate the formal expressions for the specific SDE discussed in this work.

\subsection{MZ formalism for stochastic microscopic dynamics}

The starting point for the MZ formalism is the $ N+1 $-dimensional stochastic differential equation we have introduced in Eq.~(\ref{eq:lin_eom}). Here and in the following, observables without explicit time dependence are defined as $ v_i = v_i(t=0) $. Using the backward Kolmogorov operator,
\begin{equation}\label{key}
\mathcal{K}(\bm{v}) = \sum_{i=0}^N A_{ij} v_{j} \frac{\partial }{\partial v_{i}} +\frac{1}{2} \sum_{i=0,j=0}^N \Phi_{ij} \frac{\partial^2}{\partial v_{i} \partial v_{j}},
\end{equation}
we can define a ``composition operator'' $ \mathcal{M}(t,0) = e^{t \mathcal{K}} $, which corresponds to the time-evolution operator for the $ N $-dimensional noise-averaged observable $ \tilde{\bm{u}}(t) = \mathbb{E}_{\bm{W}(t)}\left[ \bm{u}(\bm{v}(t)) | \bm{v}_0 \right] $,
\begin{equation}\label{key}
\tilde{\bm{u}}(t) = e^{t \mathcal{K}} \bm{u}(\bm{x}_0).
\end{equation}
This noise-averaged quantity $\tilde{\bm{u}}(t) $ is thus the average over all trajectories starting from the initial point $ \bm{v}_0 $ and using different realizations of the Wiener process $ \text{d} \bm{W}(t) $. The above rewriting of the SDE using the time-evolution operator $e^{t \mathcal{K}} $ is the crucial step for the application of the Mori-Zwanzig formalism, because from this point on the standard derivation can be applied by substituting the Liouville operator $ \mathcal{L}$ by the Kolmogorov operator $ \mathcal{K}$.

In the following, we will specify the observable $ u(\bm{v}(t)) = v_0(t) $ and define the Mori-Zwanzig projection operator,
\begin{equation}\label{eq:projector}
\mathcal{P}A = \frac{\langle v_0(0) A \rangle}{ \langle v_0(0)^2\rangle  } v_0(0).  
\end{equation}
The orthogonal projector is correspondingly given by $ \mathcal{Q} = 1 - \mathcal{P} $. Using the usual MZ projection formalism \cite{zhu2021effective} we can then derive the equation of motion for $ \tilde{v}_0(t) $,
\begin{equation}\label{eq:GLE_MZ}
\dot{\tilde{v}}_0(t) = -\Gamma \tilde{v}_0(t) - \int_{0}^{t} K^{\text{p}}(t-s) \tilde{v}_0(s) \text{d}s + \eta^{\text{p}}(t),
\end{equation}
with,
\begin{align}
\Gamma &= - \frac{\langle v_0(0) \mathcal{K} v_0(0) \rangle}{ \langle v_0(0)^2\rangle  } \quad &(\text{friction constant}) \label{friction_constant}\\
K^\text{p}(t) &= \frac{\langle \eta^\text{P}(t)\eta^\text{P} \rangle}{ \langle v_0(0)^2\rangle  } \quad &(\text{memory kernel})\\
\eta^\text{p}(t) &= e^{t \mathcal{Q}\mathcal{K}}\mathcal{Q}\mathcal{K} v_0(0).\quad \label{eq:noise_theo} & (\text{noise})
\end{align}
It should be noted that the second fluctuation-dissipation theorem which defines the memory kernel $ K^\text{p}(t) $ is a direct consequence of the special choice of the observable, $ u(t) = v_0(t) $, which does not have direct white noise. In the general case, however, a violation of the fluctuation dissipation theorem could be possible, as discussed in Ref.~\cite{zhu2021generalized}.  From the MZ projection operator Eq.~(\ref{eq:projector}) it can also generally be shown that the orthogonality condition 
\begin{equation}\label{eq:orthogonality}
 \langle v_0(0) \eta^\text{P}(t) \rangle = 0 
\end{equation}
is generally valid.

To interpret Eq.~(\ref{eq:GLE_MZ}) as a \emph{stochastic} differential equation we assume that the initial condition $ \bm{v}_0 $ is random and characterized by the steady-state probability distribution $ \rho_{S}(\bm{v}) $ which satisfies the Fokker-Planck equation $ \partial_t \rho_S = \mathcal{K}^* \rho_S = 0 $ for the adjoint Kolmogorov operator $ \mathcal{K}^* $. Since it can further be shown that,
\begin{equation}\label{key}
C_V(t)=\langle v_0(t) v_0(0) \rangle_{\rho_S} = \langle \tilde{v}_0(t)  \tilde{v}_0 (0) \rangle_{\rho_S},
\end{equation}
we can, in fact, interpret Eq.~(\ref{eq:GLE_MZ}) as a GLE for the dynamics of $ v_0(t) $ \cite{zhu2021effective,zhu2021generalized},
\begin{equation}\label{eq:GLE_final}
\dot{{v}}_0(t) = - \Gamma v_0(t) - \int_{0}^{t} K^{\text{p}}(t-s) {v}_0(s) \text{d}s + \eta^{\text{p}}(t).
\end{equation}
Different from the integration route, the noise in the projection route therefore purely originates from the lack of knowledge about the initial velocities of the bath particles, which is precisely the philosophy of the MZ formalism.
The memory kernel can also be directly related to the velocity autocorrelation function via the deterministic Volterra equation \cite{Zwanzig2001,Shin2010,jung2021fluctuation,zhu2021effective},
\begin{equation}\label{eq:volterra}
\dot{C}_V(t) = - \Gamma{C}_V(t)  - \int_{0}^{t} K^{\text{p}}(t-s) C_V(s) \text{d}s.
\end{equation}
This equation can immediately be derived from the general orthogonality condition Eq.~(\ref{eq:orthogonality}) and the GLE (\ref{eq:GLE_final}).

\subsection{Non-Markovian equations of motion for the colloid (projection route)}
\label{sec:theory_projection}

Specifically for the microscopic stochastic dynamics Eqs.~(\ref{eq:colloid}) and (\ref{eq:solvent}) we can explicitly write down the Kolmogorov operator $ \mathcal{K} $,
\begin{align}\label{key}
\mathcal{K} = &(-\gamma_0 v_0 + \sum_{i>0}k_i v_i) \frac{\partial}{\partial v_0} \nonumber\\
& +\sum_{i>0} (-\gamma_i v_i + b_i v_0) \frac{\partial}{\partial v_i} + T\sum_{i>0} \gamma_i \frac{\partial^2}{\partial v_i^2}.
\end{align}
The essential property of this specific Kolmogorov operator is that it transforms a linear combination of the variables $ \{ v_i \} $ into another linear combination of these variables. This property is a direct consequence of the linearity of the microscopic equations of motion. Additionally, the same holds for the projection operator. Similar to Ref.~\cite{Zwanzig2001} we can thus propose a general ansatz for the noise,
\begin{equation}\label{key}
\eta^\text{p}(t) = \sigma_0(t) v_0 + \sum_{i>0} \sigma_i(t) v_i.
\end{equation}
From the MZ projection operator formalism we further know that the noise is defined by Eq.~(\ref{eq:noise_theo})
and thus
\begin{equation}\label{eq:noise_relation}
\frac{\text{d}}{\text{d} t} \eta^\text{p}(t) = \mathcal{Q} \mathcal{K} \eta^\text{p}(t).
\end{equation}
The previous equation (\ref{eq:noise_relation}) allows us to determine the equations of motion for the time-dependent parameter $ \sigma_i(t) $ by calculating the right hand side (RHS) of Eq.~(\ref{eq:noise_relation}) explicitly. We first find,
\begin{align}
\mathcal{K} \eta^\text{p}(t) =& \left(-\gamma_0 \sigma_0(t) + \sum_{i>0} b_i \sigma_i(t) \right) v_0  \nonumber \\
&+ \sum_{i>0}  \big(k_i \sigma_0(t) - \gamma_i \sigma_i(t) \big) v_i,\label{eq:Kv0}
\end{align}
and therefore we can conclude,
\begin{align}
\mathcal{Q}\mathcal{K} \eta^\text{p}(t) =& - \left(  \sum_{i>0}\big(k_i \sigma_0(t) - \gamma_i \sigma_i(t) \big) \frac{ \langle v_0 v_i  \rangle}{\langle v_0^2 \rangle}      \right) v_0 \nonumber \\
&+ \sum_{i>0} \big(k_i \sigma_0(t) - \gamma_i \sigma_i(t) \big) v_i.
\end{align}
The RHS can be equated to the left hand side of Eq.~(\ref{eq:noise_relation}),
\begin{align}
\frac{\text{d}}{\text{d} t} \eta^\text{p}(t) =& \dot{\sigma_0} v_0 + \sum_{i>0}\dot{\sigma_i} v_i, 
\end{align}
which allows us to determine the differential equations for $ \sigma_i(t) $ via comparison of coefficients,
\begin{align}
\frac{\text{d}}{\text{d} t} \sigma_0(t) &=  - \sum_{i>0}k_i \frac{ \langle v_0 v_i  \rangle}{\langle v_0^2 \rangle}   \sigma_0(t) +  \sum_{i>0}\gamma_i \frac{ \langle v_0 v_i  \rangle}{\langle v_0^2 \rangle}   \sigma_i(t)  \nonumber\\
&= -\gamma_0 \sigma_0(t)+  \sum_{i>0} \gamma_i \frac{ \langle v_0 v_i  \rangle}{\langle v_0^2 \rangle}   \sigma_i(t),  \\
\frac{\text{d}}{\text{d} t} \sigma_i(t) &=- \gamma_i \sigma_i(t) + k_i \sigma_0(t) , \quad i>0.
\end{align}
We can further extract the initial conditions using Eq.~(\ref{eq:noise_theo}) at $ t=0 $,
\begin{equation}\label{key}
\sigma_0(0) = - \sum_{i>0} k_i \frac{ \langle v_0 v_i  \rangle}{\langle v_0^2 \rangle} = -\gamma_0 , \quad \sigma_i(0) = k_i.
\end{equation}
The former equation for $ \sigma_0(0) $ must hold because of Eq.~(\ref{eq:rel_v01}).

In situations, where $ \langle v_j v_i  \rangle = 0 $, $ i\neq j $ (i.e., at equilibrium), these differential equations have the trivial solution,
\begin{equation}\label{key}
\sigma_0(t) = 0, \quad \sigma_i(t) = k_i \exp(- \gamma_i t)\quad i>0,
\end{equation}
and thus reduce to the relation $	\langle \eta^\text{p}(t) \eta^\text{p}(0) \rangle = \sum_{i>0} k_i^2  \exp (-\gamma_i t)$ as obtained by the integration route.  Additionally, in the non-equilibrium situation where $ \gamma_0=0 $ and $ \gamma_i=\gamma \nobreakspace \forall i $, it can be directly inferred that $\sigma_0(t) = 0$. Therefore, also in this case the projection route and the integration route are equivalent and the 2FDT is always fulfilled. 

More generally, however, $ \sigma_0(t) $ will be different from zero and thus the projection route will yield different results than the integration route. In these cases, the linear equations of motion for $ [\bm{\sigma}(t)]_i = \sigma_i(t)\nobreakspace i=0,\ldots, N $ can be written as $ \dot{\bm{\sigma}} = -\bm{\gamma} \bm{\sigma}  $, with the friction matrix,
\begin{equation}\label{key}
\bm{\gamma} = \left(\begin{array}{c|c} 
\gamma_0 &  \sum_{i=1}^N \gamma_i \frac{ \langle v_0 v_i  \rangle}{\langle v_0^2 \rangle} \hat{e}_i^T\\
\hline
\sum_{i=1}^{{N}} k_i \hat{e}_i & \gamma_i \delta_{ij}
\end{array} \right),
\end{equation}
where $ \hat{e}_i  $ is the unit vector with entry $ 1 $ at position $ i $ and 0 otherwise.
The solution of the above linear equations is given by the matrix equation $ \bm{\sigma}(t) = \exp(-\bm{\gamma} t) \bm{\sigma}(0), $ which can be easily evaluated numerically. The final autocorrelation function for the noise can then be calculated as,
\begin{equation}\label{eq:sol_MZ_eta}
C_\eta^\text{p}(t)=\langle \eta^\text{p}(t) \eta^\text{p}(0) \rangle = \sum_{i,j} \sigma_i(t) \sigma_j(0) \langle v_i v_j \rangle.
\end{equation}
Since the 2FDT is strictly fulfilled for the MZ formalism \cite{zhu2021generalized}, the memory kernel can be immediately inferred via,
\begin{equation}\label{eq:2FDT_MZ}
K^\text{p}(t)= \langle v_0^2 \rangle^{-1} C_\eta^\text{p}(t).
\end{equation}

With the same procedure we can determine the velocity autocorrelation function from the projection operator formalism,
\begin{align}
v(t) &= e^{\mathcal{K}t} v_0\\
\Rightarrow \frac{\text{d}}{\text{d} t}v(t) &= \mathcal{K} v_0. \label{eq:cv_ansatz}
\end{align}
Here, we use the ansatz $v(t) = \alpha_0(t) v_0 + \sum_{i>0}\alpha_i(t) v_i$ to find $\mathcal{K} v(t)$ in similar spirit as Eq.~(\ref{eq:Kv0}). By inserting $ \mathcal{K} v(t) $ into Eq.~(\ref{friction_constant}) we can immediately conclude that $ \Gamma=0 $ for our model. Combining Eq.~(\ref{eq:cv_ansatz}) with the explicit expression for $ \mathcal{K} v(t) $ we can infer the equations of motion for time-dependent parameters $ \alpha_i(t) $,
\begin{align}
\frac{\text{d}}{\text{d} t} \alpha_0(t) &= -\gamma_0 \alpha_0(t) + \sum_{i>0} b_i \alpha_i(t)  \\
\frac{\text{d}}{\text{d} t}\alpha_i(t) &=  - \gamma_i \alpha_i(t) + k_i \alpha_0(t), \quad i>0,
\end{align}
with initial conditions,
\begin{equation}\label{key}
\alpha_0(0) = 1, \quad \alpha_i(0) = 0.
\end{equation}
Finally, we can calculate the velocity autocorrelation function,
\begin{equation}\label{eq:sol_MZ_V}
C_V(t) = \sum_{i} \alpha_i(t) \langle v_i v_0 \rangle.
\end{equation}

Interestingly, the dynamics of $ \alpha_i(t) $ and $ \sigma_i(t) $ are very similar and their equations of motion only differ by one term. The former corresponds to the ``real'' dynamics, given by the usual time evolution operator, $ e^{t \mathcal{K}} $, and the latter to the ``orthogonal'' dynamics, $ e^{t \mathcal{Q} \mathcal{K}} $.

\section{Numerical results}
\label{sec:results}

We will present results for four different models: The first corresponds to equilibrium dissipative dynamics, the second to a non-equilibrium system where both solutions fulfill the 2FDT. The third model has similarities to a non-equilibrium system with time-delayed feedback control \cite{loos2014delay,loos2019fokker,Loos2020,e23060696}. All these models consist of a colloid coupling to two bath particles, $ N=2 $. The fourth model was used to understand persistent active motion \cite{wu2000particle,PhysRevLett.117.038103} and only contains a single bath particle coupling to the colloid, $ N=1 $.

To supplement and illustrate the above theoretical results we will also perform computer simulations of the microscopic stochastic model, Eqs.~(\ref{eq:colloid}) and (\ref{eq:solvent}). Details for the simulations can be found in Appendix~\ref{sec:integrator}.

\subsection{From equilibrium to non-equilibrium}

\begin{figure}
		\hspace*{-0.5cm}\includegraphics[scale=1]{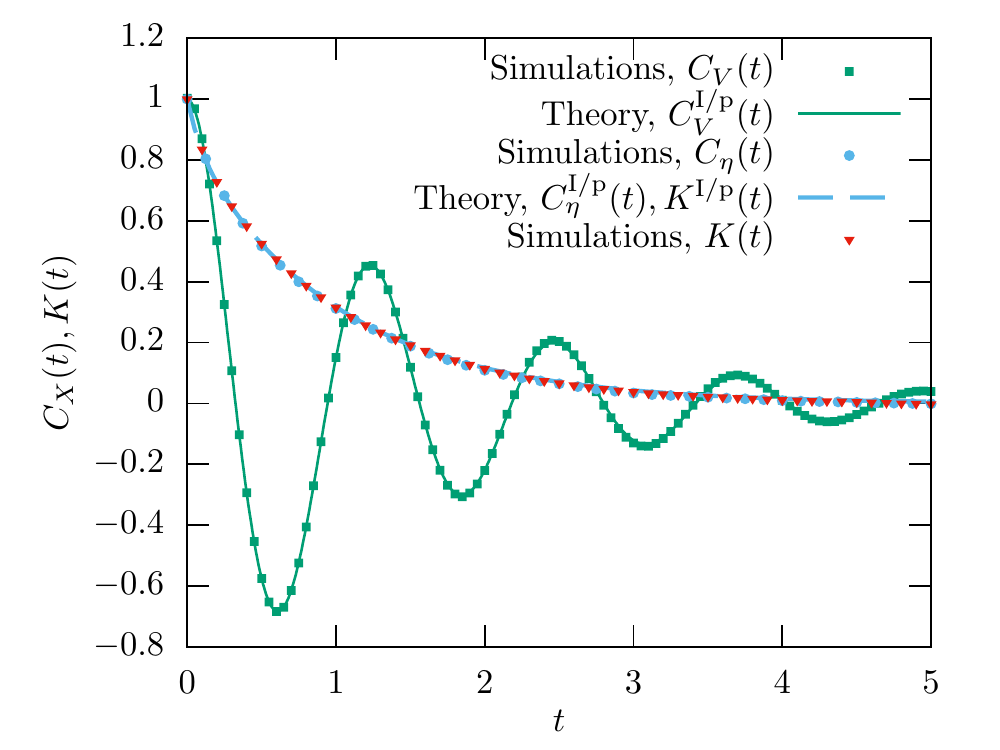}
	\caption{Velocity and noise autocorrelation for the equilibrium system with parameters $ k_1=5 $, $ k_2=2 $ and $ \gamma_2=10 $ (EQ). The correlation functions $ C^\text{p} $ were determined based on the analytical results Eqs.~(\ref{eq:sol_MZ_eta}) and (\ref{eq:sol_MZ_V}), and $ C_\eta^\text{I} $ from Eq.~(\ref{eq:noise_integrate}).  Shown is also the memory kernel, $ K^\text{I/p} $, as calculated from Eqs.~(\ref{eq:memory_integrate}) and (\ref{eq:2FDT_MZ}). $ C_V^\text{I} $ is based on the analytical expressions derived in Section~3.1. of Ref.~\cite{doerries2021correlation}. }
	\label{fig:eq}
\end{figure}

Within our model an equilibrium system is given by a vanishing instantaneous friction, $ \gamma_0 = 0 $, and reciprocal interactions, $ k_i = -b_i. $ In equilibrium, all cross-correlations vanish, $ E_{ij} = 0,\, i\neq j $, and the diagonal components are given by $ E_{ii} = k_B T $ (see Table~\ref{tab:inst_fluct}). The time autocorrelation function of the noise, $ C_\eta(t) $ is equivalent between the projection route and the integration route, as is shown in Fig.~\ref{fig:eq}, and the 2FDT is strictly fulfilled.

\renewcommand{\arraystretch}{1.2}
\begin{table}
	\centering \begin{tabular}{ccccccc} 
		model & $E_{00}$  & $E_{01}$ & $E_{02}$  & $E_{11}$ & $E_{12}$  & $E_{22}$ \\
		EQ & $1.0$ &  $0.0 $ & $0.0$ & $ 1.0$ & $0.0$ & $1.0$  \\
				\myrowcolour
		NEQ1 & $1.38$ &  $-0.165 $ & $0.41$ & $ 1.83$ & $-1.20$ & $1.83$  \\

		NEQ2 & $ 2.46 $  & $ -0.41 $   & $1.39 $ & $ 3.03 $&$ -0.82 $&$ 1.70 $ \\
		\myrowcolour
		NEQ3& $ 0.83 $  & $ 0.83 $   & $ - $ & $ 1.0 $& $ - $&$ - $\\ 
	\end{tabular} 
	\caption{Numerical values for the instantaneous fluctuations of the different models studied in this work, as calculated from Eqs.~(\ref{eq:inst_fluct}). The model parameters are defined in Fig.~\ref{fig:eq} (EQ), Fig.~\ref{fig:neq1} (NEQ1), Fig.~\ref{fig:neq2} (NEQ2) and Fig.~\ref{fig:neq3} (NEQ3), respectively.}
	\label{tab:inst_fluct}
\end{table} 
\renewcommand{\arraystretch}{1.0}

\begin{figure}
	\hspace*{-0.75cm}\includegraphics[scale=1]{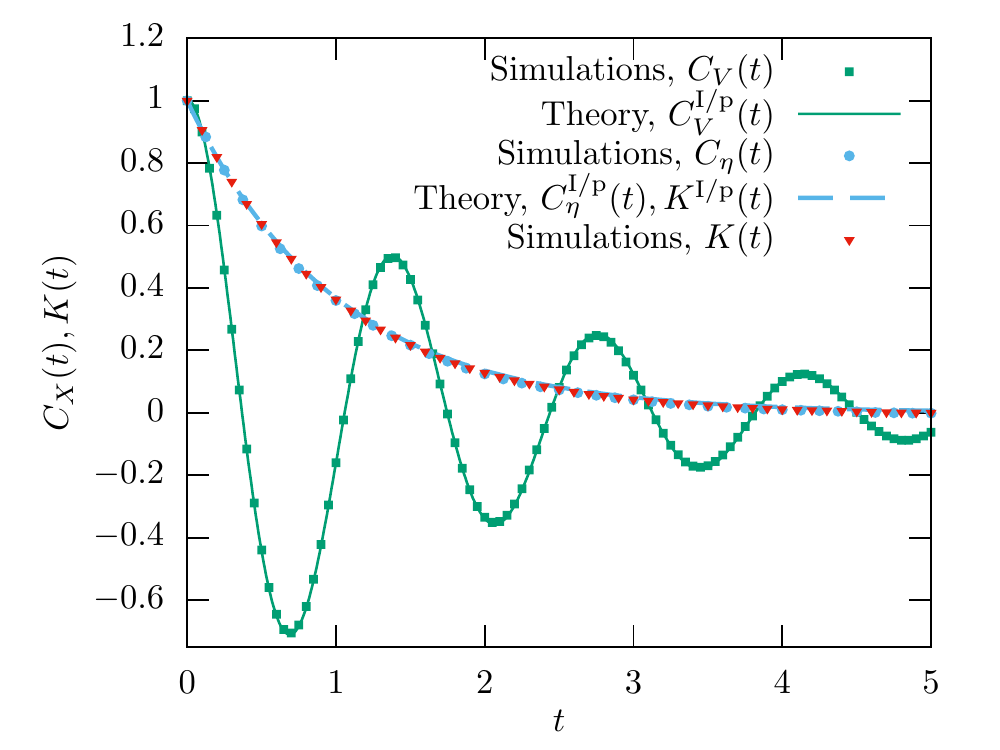}
	\caption{Velocity and noise autocorrelation for the non-equilibrium system with parameter $ k_1=5=-b_1 $, $ k_2=2=b_2 $, $ \gamma_0=0 $ and $ \gamma_2=1 $ (NEQ1). The theoretical curves are calculated using the same relations as in Fig.~\ref{fig:eq}. }
	\label{fig:neq1}
\end{figure}

When introducing non-reciprocal interactions between the colloid and the bath particles by setting $ b_2 = k_2 $, the system becomes non-equilibrium but attains a steady state \cite{zhu2021effective,doerries2021correlation}. In Fig.~\ref{fig:neq1} results are shown for the special case in which $ \gamma_1 = \gamma_2. $ As has been discussed theoretically in the previous sections, despite being far from equilibrium, the 2FDT is fulfilled also by the solution via the integration route. 

The results for both equilibrium and non-equilibrium are in very good agreement with computer simulations. We therefore have consistent results for a specific non-equilibrium model in which the two theoretical routes lead to identical solutions.



\subsection{Non-equilibrium system: time-delayed feedback control}
\label{sec:feedback}

\begin{figure}
	\hspace*{-0.75cm}\includegraphics[scale=1]{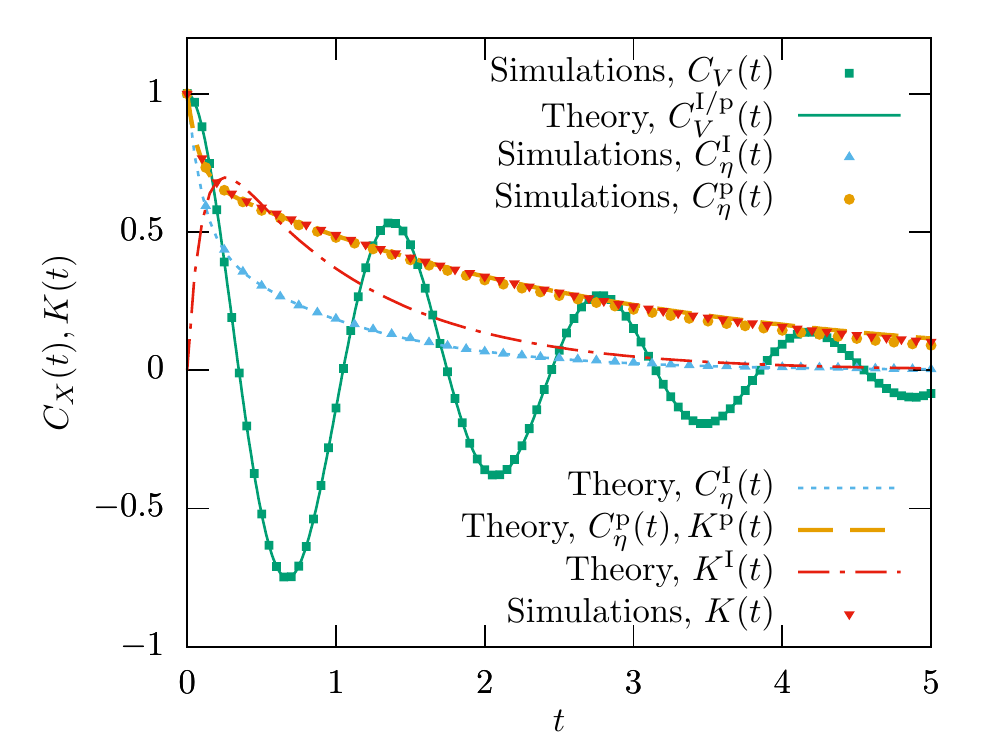}
	\caption{Velocity and noise autocorrelation for the time-delayed feedback control system with parameters $ k_1=5=-b_1 $, $ k_2=5=b_2 $, $ \gamma_0=2 $, $ \gamma_2=10 $ (NEQ2). The theoretical curves are calculated using the same relations as in Fig.~\ref{fig:eq}. The numerical values for $ C_V^\text{p}(t) $ and $ C_V^\text{I}(t) $ are identical and thus not shown explicitly. }
	\label{fig:neq2}
\end{figure}

The situation changes when studying more general non-equilibrium systems. First, we investigate a model which shows a maximum in the memory kernel $ K^\text{I}(t) $ (see red, dashed-dotted line in Fig.~\ref{fig:neq2}). Such models have been discussed before in the context of stochastic thermodynamics, where an entity (in this case the colloid) shows a marginal response to its current state but exhibits a much more pronounced time-delayed ``feedback'' \cite{e23060696}. Different from the memory kernel $ K^\text{I}(t) $ we find that the autocorrelation function of the noise, $ C_\eta^\text{I}(t) $, decays monotonically and thus the 2FDT is violated. The latter must be the case for any memory kernels for which a time $ t^* $ can be found with $ K(t^*) > K(0) $, since $ C_\eta^\text{I}(t) $ is a correlation function and thus fulfills the relation $ |C_\eta^\text{I}(t^*)| < C_\eta^\text{I}(0),\,\, \forall t^* $ \cite{forster2018hydrodynamic,franosch2014long}. 

 When applying the Mori-Zwanzig projection operator formalism to the same system we observe that the traces of this time-delayed mechanism completely vanish and the memory kernel $ K^\text{p}(t) $ decays monotonically in time (see red triangles in Fig.~\ref{fig:neq2}). Different from the integration route, however, the result from the projection operator formalism fulfills the 2FDT. In this example, the relaxation time of the correlation function $ C^\text{p}_\eta(t) $ is much larger than the one of $ C^\text{I}_\eta(t) $, showing that both qualitative and quantitative features are different between the projection route and the integration route.
 
  It is important to note that although the memory kernels themselves differ significantly, the distinct noise correlations then finally lead to the same velocity autocorrelation functions $C_V(t)$, as shown in Fig.~\ref{fig:neq2}.

\subsection{Non-equilibrium system: active bacteria bath}
\label{bacteria}

\begin{figure}
	\hspace*{-0.75cm}\includegraphics[scale=1]{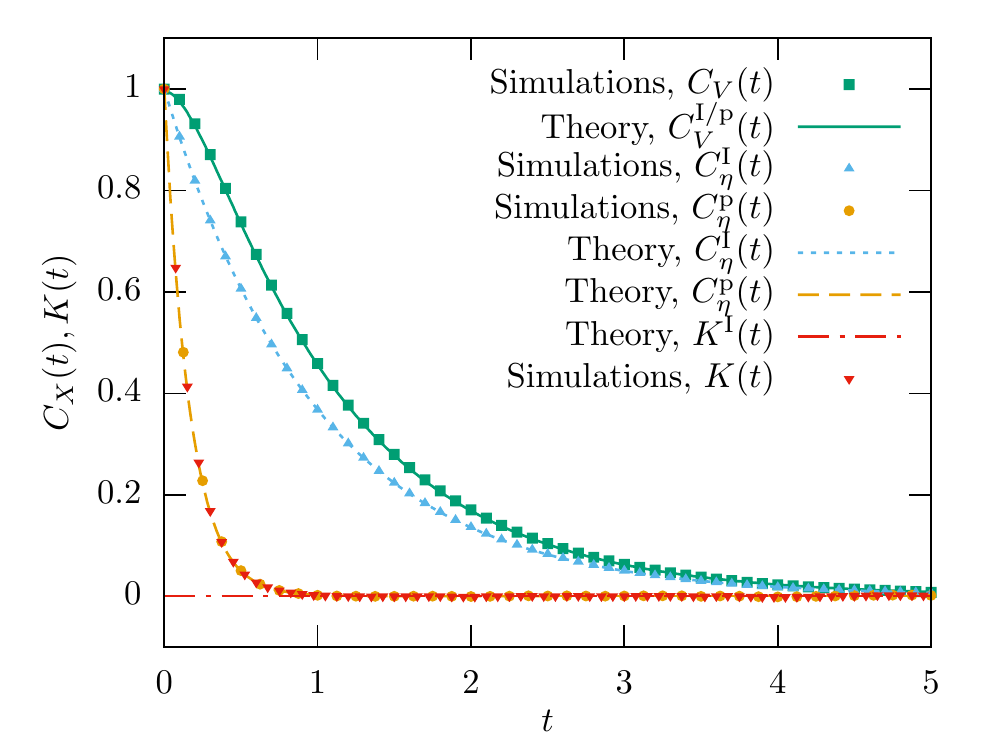}
	\caption{Velocity and noise autocorrelation for the active bacteria bath model with parameters, $ k_1=5 $, $ b_1=0 $ and $ \gamma_0=5 $ (NEQ3). The theoretical curves are calculated using the same relations as in Fig.~\ref{fig:eq}. The numerical values for $ C_V^\text{p}(t) $ and $ C_V^\text{I}(t) $ are identical and thus not shown explicitly.}
	\label{fig:neq3}
\end{figure}

The last example we study includes only one bath particle and is equivalent to the model suggested by Wu and Libchaber \cite{wu2000particle}. The bath particle is supposed to represent an active bacteria bath which couples collectively and leads to non-equilibrium persistent motion of the passive colloid. The relaxation time of the bath is connected to the orientational diffusion of the bacteria \cite{wu2000particle}. The fluctuations are thus not thermal in nature but arise from the random reorientations of the bacteria, which themselves do not couple to the colloid, $ b_1=0 $.  Due to the missing coupling of the bath to the colloid, the memory kernel in our model is simply $ K^\text{I}(t) = 0, $ thus all non-trivial dynamics arise from the time-correlation function of the noise, which decays exponentially. In recent years the model has also been intensively studied as model for active particles themselves called the active Ornstein-Uhlenbeck particles \cite{PhysRevLett.117.038103}.

As in the previous example, the decay of the memory kernel and the noise as calculated via the projection route is very different from that via the integration route and fulfills the 2FDT. Different from the previous case, however, in this case the noise autocorrelation function of the projection route, $ C_\eta^\text{p}(t) $, decays much faster than $ C_\eta^\text{I}(t) $.

We have also investigated intensively the statistics of the reconstructed noise from the simulations for both the integration route and the projection route similar to Refs.~\cite{Shin2010,jung2021fluctuation} The probability distribution for the noise was always Gaussian, which is a direct consequence of the Gaussian white noise acting on the bath particles as defined in Eq.~(\ref{eq:solvent}).

\section{Conclusions and Outlook}
\label{sec:conclusions}

In this work we have proven for a linear stochastic model that in various non-equilibrium conditions coarse-graining via direct integration and via the projection operator formalism leads to distinct results. Our purely analytical calculations have been supplemented by numerical calculations for different previously studied non-equilibrium models, showing that the two solutions differ quantitatively and qualitatively and can even lead to memory and noise autocorrelations on very different timescales. Despite these differences, the resulting velocity autocorrelation functions will, by construction, be identical between the two models.

A complementary argumentation to the explicit calculations performed in this work is based on the theory of correlation functions \cite{forster2018hydrodynamic,franosch2014long}. The validity of the 2FDT inherently requires the memory kernel to be a correlation function because it is identical to the noise autocorrelation function (multiplied by a constant). Consequently, any memory kernel determined via the integration route which does not fulfill the properties of a correlation function (i.e., positive spectrum) must automatically be different from the projection route.

An important question that arises immediately from the results shown in this work is at  which type of non-equilibrium the projection route and the integration route start to differ. Two conclusions that can be drawn from the present work is that non-equilibrium is a necessary but certainly not a sufficient condition. From the example in Section \ref{bacteria} for the bacteria baths it can also be concluded that systems exists in which $ K^\text{I}(t) $ is a correlation function but the 2FDT is still violated and thus differences exist between the projection route and the integration route. Within the present toy model it appears that the 2FDT will be violated by the exact solution for any non-equilibrium system with non-reciprocal interactions and $ \gamma_1 \neq \gamma_2 $.

We strongly expect that the conclusions drawn here will similarly apply to more complex, atomistic systems, leading to the fundamental question which description to choose as coarse-grained model in non-equilibrium situations. On the one hand, the memory kernel $ K^\text{I}(t) $ describes the exact solution of the stochastic microscopic model, however, it cannot be constructed straightforwardly for more complex systems in which analytical results are not achievable.  On the other hand, the solution of the MZ projection operator formalism represents a systematic way of calculating coarse-grained models which fulfill the 2FDT and also certain orthogonality constraints between the noise and the coarse-grained particles. One consequence of this constraint is the deterministic Volterra equation (\ref{eq:volterra}) which is of fundamental importance for many theoretical models, for example in mode-coupling theory \cite{Bengtzelius_1984}. However, also the MZ projection operator formalism does not make any predictions about the statistics of the noise beyond the 2FDT.

More generally phrased, the results from this paper and also Refs.~\cite{maes2014second,netz2018fluctuation} indicate that coarse-graining in non-equilibrium has to go beyond the dimensional reduction via time autocorrelation functions only.  One possible numerical route to find alternative memory kernels has been suggested in Ref.~\cite{netz2018fluctuation} via additional measurement of the positional response function to an external force. It could also be possible to construct a projection operator formalism which is not based on the MZ type of projector as defined in Eq.~(\ref{eq:projector}) but via more general projectors, as in Refs.~\cite{Zwanzig1961,grabert2006projection,10.1209/0295-5075/ac35ba}. The immediate question to ask would be: Is it possible to formulate a projection operator $ \mathcal{P} $, which leads to the linear GLE (\ref{eq:GLE_integrate}) with the memory kernel $ K^\text{I}(t) $? A related question comprises the entropy production in such coarse-grained systems and whether it is possible to connect thermodynamic quantities such as entropy \cite{PhysRevE.101.022120,Loos2020,e23060696} as discussed intensively in the framework of stochastic thermodynamics \cite{Speck2016,PhysRevLett.119.258001,fodor2021irreversibility} to dynamic coarse-graining, for example via projection operator techniques? We believe that answering these questions would pave the way towards systematic dynamic coarse-graining of non-equilibrium soft matter systems with applications to active microrheology \cite{Puertas_2014,jung2021fluctuation}, active matter \cite{Maes2014,Speck2016,CHAKI2019121574}, but also non-stationary situations such as crystallization \cite{kuhnhold2019derivation}, colloid self-assembly \cite{self_assembly2010} or at phase transitions \cite{PhysRevE.103.022102} using a non-stationary GLE \cite{doi:10.1063/1.5006980,doi:10.1063/5.0049693}.

\section*{Acknowledgements}

The author thanks Yuanran Zhu, Thomas Franosch, Friederike Schmid, Bernd Jung, Sabine Klapp, Timo Dörries and Sarah Loos for helpful discussions and critical reading of the manuscript, as well as Bernd Jung and Timo Dörries for providing an early version of the simulation code and the Mathematica script to evaluate the expressions derived in Section 3.1. of Ref.~\cite{doerries2021correlation}, respectively.

This work was partly supported by a short term fellowship (PE21004) from the Japan Society for the Promotion of Science (JSPS).
  
\FloatBarrier

\appendix

\section{Computer simulations}
\label{sec:integrator}

To supplement and illustrate the above theoretical results we will also perform computer simulations of the microscopic stochastic model, Eqs.~(\ref{eq:colloid}) and (\ref{eq:solvent}). In recent years several ways have been suggested to discretize and integrate such kind of equations of motion \cite{ceriotti2010colored,baczewski2013numerical,leimkuhler2020efficient,duong2021accurate}. Here, we will use an approach similar to Ref.~\cite{gronbech2013simple} to derive the integrator with timestep $ \Delta t $ for the velocities at time $ v^n_i = v_i(t=n\Delta t) $. As only assumption for the derivation of the integrator we approximate,
\begin{equation}\label{key}
\int_{t}^{t+\Delta t} v_i(t) \text{d} t \approx \frac{\Delta t}{2} \left\langle v_i^{n+1} + v^n_i\right\rangle,
\end{equation}
which introduces errors that scale as $\mathcal{O}(\Delta t^3)$ \cite{gronbech2013simple}. Based on this relation we can integrate the SDEs (\ref{eq:colloid}) and (\ref{eq:solvent}) in time between $ t $ and $ t + \Delta t $ to find (without introducing further approximations),
\begin{align}
v_0^{n+1} &= v_0^n - \gamma_0 \frac{\Delta t}{2} (v_0^{n+1} + v_0^n) + \sum_i k_i \frac{\Delta t}{2} (v_i^{n+1}+v_i^{n}) \label{eq:integratev0}\\
v_i^{n+1}  &= v_i^n -\frac{\gamma_i \Delta t}{2} (v_i^{n+1}+v_i^{n}) + \frac{b_i \Delta t}{2} (v_0^{n+1} + v_0^n) + \xi_i^{n} \label{eq:integratevi}
\end{align}
where $\langle \xi_i^{n}\xi_j^{n+k}\rangle = 2 k_B T \gamma_i \Delta t  \delta_{ij} \delta_{k0}  $. The previous equation (\ref{eq:integratevi}) can then be resolved for $v_i^{n+1}$,
\begin{align}
v_i^{n+1} (1 + \frac{\gamma_i \Delta t}{2})  &= v_i^n (1 - \frac{\gamma_i \Delta t}{2}) + \frac{b_i \Delta t}{2} (v_0^{n+1} + v_0^n) + \xi_i^{n} \nonumber\\
v_i^{n+1}  &= e_i v_i^n  + \frac{c_i b_i \Delta t}{2} (v_0^{n+1} + v_0^n) + c_i \xi_i^{n}, \label{eq:vin}
\end{align}
with $ c_i = \left(1+\frac{\gamma_i \Delta t}{2}\right)^{-1} $, $ e_i = c_i \left( 1-\frac{\gamma_i \Delta t}{2}\right) $. Inserting Eq.~(\ref{eq:vin}) into Eq.~(\ref{eq:integratev0}) and resolving for $ v^{n+1}_0 $ then leads to the final integrator,
\begin{align}
v_0^{n+1}  &= \frac{e_0 + f}{1 - f} v^n + \frac{1}{1- f} \sum_i g_i (2 y_i^n  +  \xi_i^{n}) + \frac{c_0 \xi_0^{n}}{1 - f}, \nonumber\\
v_i^{n+1}  &= e_i y_i^n  + \frac{c_i b_i \Delta t}{2} (v^{n+1} + v^n) + c_i \xi_i^{n}. \label{eq:integrator}
\end{align}
Here, we have defined the integration constants, $ f = c_0 \sum_i k_i \frac{\Delta t^2}{4} c_i b_i  $ and $ g_i = c_0 \frac{ k_i c_i \Delta t}{2}   $.

For all simulations we choose a timestep of $ \Delta t = 0.005 $ and equilibrate the system for $ 10^5 $ steps before evaluating the trajectory for at least $ 10^7 $ steps. From these simulation results, we determine the instantaneous fluctuations $ E_{ij} $ and the velocity autocorrelation function $ C_V(t) $ in the steady state. From the simulations, we calculate the memory kernel $ K^\text{p}(t) $ via numerical inversion of the Volterra integral, Eq.~(\ref{eq:volterra}), similar to Ref.~\cite{Shin2010}. Having extracted $ K^\text{p}(t) $ we can directly calculate $ \eta^\text{p}(t) $ from the simulation trajectories as the only unknown quantity in Eq.~(\ref{eq:GLE_final}). Similarly, we extract $ \eta^\text{I}(t) $ by using the theoretically calculated $ K^\text{I}(t) $ from Eq.~(\ref{eq:memory_integrate}).

\bibliography{library_local.bib}

\end{document}